# A topological extension of GR: Black holes induce dark energy

**M Spaans**[1]

Kapteyn Astronomical Institute, University of Groningen, 9700AV Groningen, The Netherlands

E-mail: spaans@astro.rug.nl

**Abstract.** A topological extension of general relativity is presented. The superposition principle of quantum mechanics, as formulated by the Feynman path integral, is taken as a starting point. It is argued that the trajectories that enter this path integral are distinct, despite any quantum uncertainty in geometry, and thus that space-time topology is multiply connected. Specifically, space-time at the Planck scale consists of a lattice of three-tori that facilitates many distinct paths for particles to travel along. To add gravity, mini black holes are attached to this lattice. These mini black holes represent Wheeler's quantum foam and result from the fact that GR is not conformally invariant. The stable creation of such mini black holes is found to be caused by the existence of macroscopic (so long-lived) black holes. This connection, by which macroscopic black holes induce mini black holes, is a topological expression of Mach's principle. The proposed topological extension of GR can be tested because, if correct, the dark energy density of the universe should be linearly proportional to the total number of macroscopic black holes in the universe at any time. This prediction, although strange, agrees with current astrophysical observations.

## 1. Quantum mechanics, gravity and topology

The success of quantum mechanics and Einstein gravity are overwhelming. Hence, one should take great care in preserving the beautiful features of both, while pursuing the mystery of dark energy. A topological extension of GR, which allows for quantum aspects, is discussed here, following [1].

The Feynman path integral provides a very elegant formulation of quantum mechanics. In it, the motion of matter, like a particle world line, derives from a superposition of wave amplitudes along many different paths. To physically construct a world line it thus seems reasonable to assume that nature knows about all the different paths that particles can follow. After all, when one takes paths as the central objects in quantum mechanics, then one must have a means to identify them, whatever the erratic trajectories that Planck-scale fluctuations in geometry allow. Indeed, many of the paths that enter the Feynman path integral are continuous, but not necessarily differentiable. To distinguish such paths, so despite any Planck-scale quantum uncertainty in geometry, one can use topology because it allows for the continuous deformation of space-time. Therefore, one can construct distinct paths through space-time by assigning a global multiply connected character to space-time. Specifically, [1] proposes a lattice of three-tori embedded in 4 dimensions to create many topologically distinct paths between any two space-time points, and thus facilitate the Feynman path integral, as follows. A three-torus is a solid cube with its faces identified through the 4th dimension. A lattice of three-tori can be constructed by connecting each face of a three-torus to a face of a neighbouring three-torus through a 3D junction. Wave amplitudes can then freely travel along the lattice junctions and pick up phase shifts along the 3 loops comprising any three-torus, $T^3 = S^1 \times S^1 \times S^1$. The local isotropy and flatness of the three-torus allows one to define relativistic quantum mechanics in small space-time patches, in the usual manner, while the global multiply connected nature of the lattice of three-tori implements the superposition principle using space-time itself.

---

[1] To whom any correspondence should be addressed.

The Einstein equation has many symmetries, but it is not invariant under local conformal (scale changing) transformations. As a result, one finds large metric fluctuations when one goes to small spatial scales. In fact, at the Planck scale of $l_P \sim 10^{-33}$ cm, these fluctuations in space-time become huge (even singular) and space-time should take the form of a quantum foam with Planck mass ($m_P \sim 10^{-5}$ g) black holes (BHs), popping out off and into the vacuum [2]. Space-time thus enjoys the presence of mini BHs at the Planck scale, i.e., is also multiply connected when one considers Einstein gravity at the Planck scale. A fundamental problem with the quantum foam is that it fluctuates violently on a Planck time of $t_P \sim 10^{-43.5}$ sec and is therefore difficult to conceive as a stable structure. Nevertheless, as a step towards a theory of quantum gravity, and using the freedom to choose the (global) topology of space-time within GR, one can think of a Planck scale space-time that comprises a lattice of three-tori with an unstable foam of mini BHs attached to it: Use topology to identify paths despite any quantum uncertainty in geometry. However, this model is quite heuristic and without dynamical content. This can be improved as follows.

One first notes that any topological space-time manifold can be constructed by adding together three-dimensional building blocks called prime manifolds. These primes (e.g., handles, three-tori, three-spheres, etc.) lend their name from the property that they themselves cannot be written as the connected sum of smaller three-dimensional units. Furthermore, interactions among those primes can never change their intrinsic properties, like the number of closed loops they contain. Finally, following the superposition principle of quantum mechanics, any property of Planck scale topology must be carried by the linear sum of these primes. One can then conclude that a plausible linear basis set consists of the non-chiral prime manifolds $S^1 \times S^2$, $S^3$ and $T^3$, where 1) the non-irreducible handle manifold appears as part of the large scale solution to the Einstein equation, i.e. the Schwarzschild BH solution; 2) the topologically trivial three-sphere is a unit element required for the large scale vacuum limit; 3) the three-torus is confined to the Planck scale and implements the superposition principle.

To implement topology driven dynamics on this basis set of primes, one introduces a loop algebra with operators that create (T*) and destroy (T) the $S^1$ loops contained in a prime P. The central topological object here is the loop because quantum mechanical superposition deals with paths and any pair of distinct paths, with the same starting and end point, is homeomorphic to a loop. Similarly, the presence of a handle $S^1 \times S^2$ requires a loop and allows a pair of paths in the form of a loop that has one branch going though the handle and the other around it. One can define topological dynamics through

$$T^*P = P \times S^1 \quad \text{and} \quad TP = nQ, \qquad (1)$$

with n the number of closed loops in P while Q denotes P with a loop contracted to a point. Details on the loop algebraic equations of motion are provided in [1], but it is sufficient to realize that the action of the scalar operator N=(TT*+T*T) amounts to

$$NT^3 = 7T^3, \quad NS^1 \times S^2 = 3S^1 \times S^2 \quad \text{and} \quad NS^3 = S^3. \quad (2)$$

Hence, the operator N defines the multiplicity of the primes and as such allows one to build a global topological manifold with heptaplets, triplets and singlets of the primes $T^3$, $S^1 \times S^2$ and $S^3$, respectively, under the repeated action of N. That is, a single BH (three-torus) *induces* a triplet (heptaplet) of BHs (three-tori). For the Planck scale, one can thus construct a lattice of three-tori, $L(T^3)$, but BHs carry mass and are subject to Hawking evaporation. Therefore, N induces triplets of BHs with a success that depends on the longevity of existing BHs, and this allows for topology change.

The quantum foam stability problem that plagues GR at the Planck scale is then solved automatically since macroscopic (so long lived) BHs induce extra pairs of mini BHs every Planck time under the action of the construction operator N. Macroscopic BHs are topologically indistinguishable from mini BHs, but are stable against Hawking evaporation, unlike their mini BH counterparts. Hence, an ensemble of macroscopic BHs provides a naturally stable quantum foam. Furthermore, topological induction of mini BHs extends Mach's principle: Einstein's interpretation dictates that the global distribution of matter determines the (changes in) local geometry, and vice versa. Wheeler's local quantum foam, with its non-trivial topology in the form of mini BHs, now attached to the lattice of three-tori, is a quantum expression of GR. Therefore, following Einstein's ideas for geometry, matter that is globally in the form of macroscopic BHs should determine the (changes in) Planck scale topology, and vice versa [1].

The spectrum of the operator N is discrete, as one would expect because topology, with its invariance under continuous deformations of space-time, cares only about countable quantities (topological invariants). Hence, the number of macroscopic BHs (*not* their space density) in the entire universe, a global quantity, determines the local occurrence of mini BHs. To be more quantitative about this, one must compute with what probability $\delta$ Wheeler's quantum foam enjoys a mini BH in every Planckian volume, given a specific number of macroscopic BHs. The topological extension of Mach's principle then states that the total number of macroscopic BHs in the universe determines the properties of the quantum foam on $L(T^3)$. Hence, $\delta$ is *linearly* proportional to the number of macroscopic BHs in the entire three-volume of the universe, for any given time slice of thickness $t_P$.

Because topology allows one to identify all space-time points through the underlying lattice of three-tori, violations in causality do not occur as nature acquires information on the total number of macroscopic BHs in the entire universe, so in all of three-space, for every time slice. The induced mini BHs constitute a dark energy density because their creation as three-dimensional objects requires an increase in four-volume to embed them. In all, the dark energy density, $\Lambda(z)$, is linearly proportional to the total number of macroscopic BHs in the entire three-volume of the universe, $N_{BH}(z)$, at a given redshift z. In this, redshift is a distinctive measure of time, in the usual 3+1 decomposition for the evolution of the three-geometry and three-topology of space-time. I.e., redshift labels different global time slices in the geometrical and topological dynamics of three-space. Algebraically,

$$\Lambda(z) = \delta(z)\, m_P / l_P^3, \quad (3)$$

with

$$\delta(z) = N_{BH}(z)\, l_P^3 / l_I^3, \quad (4)$$

and $l_I$ is the size of the universe when the first BH forms that exists for longer than a Hubble time. I.e., $l_I$ is *frozen in* when the quantum foam stabilizes and one can speak of a universe with a specific global topology in the sense of Mach. Conversely, before macroscopic BHs exist, inflation can be driven by an exponentially fast growing number of mini BHs, see [1] for details. Each mini BH pair, which for an external observer seems to appear spontaneously in some region of the quantum foam, contributes about $m_P$ gram of mass per Planckian volume. Equation (4) shows that, for $N_{BH}$ independent of redshift, any local observer concludes that the dark energy density $\Lambda$ in g cm$^{-3}$ remains constant as the universe expands. In all, one has the local space-time quantum foam that Wheeler envisioned, but globally constrained by the total number of macroscopic BHs in the entire universe.

## 2. Testing the Connection between Dark Energy and Macroscopic Black Holes

As discussed above, the dark energy density of our universe is linearly proportional to the total number of macroscopic BHs in three-space at any redshift. Very practically, this means that if the number of BHs in the entire universe at redshift 1 is twice higher than at redshift 2, then Λ in g cm$^{-3}$ doubles from z=2 to z=1. In [3] a detailed calculation on the time dependence of the dark energy density is presented. Here, a simple estimate is provided. The bulk of the stars, and thus stellar BHs, in the universe appears to be present as early as z=1, when a $(1+z)^4$ decline ensues in the cosmic star formation rate [4]. Hence, since BH formation follows high-mass (>8 $M_o$) star formation, one expects to find an effectively constant Λ for z<1, when the bulk of the BHs in the universe is in place. Conversely, given the rapidity with which massive stars/BHs are produced around z=1-3, this epoch should exhibit a strong decrease in the number of BHs, and thus Λ, from z=1 to z=3.

These qualitative expectations can be quantified as follows. For a local observer now, it is natural to consider Λ(z) normalized to its present value Λ(0), where the latter corresponds to the current number of BHs in the observable universe. The ratio Λ(z)/Λ(0) can now be expressed analytically, in terms of $dN_{BH}/dt$ for observers that measure the rate of macroscopic BH formation in their comoving volume, as

$$\Lambda(z)/\Lambda(0) = \int_y^z dt(z)\, dN_{BH}/dt \, / \int_y^0 dt(z)\, dN_{BH}/dt \approx (1+z)^{-0.36} \text{ for } z<1, \quad (5)$$

for a flat WMAP7 cosmology, BH formation starting at some large redshift y (~30-1000) but at a very modest pace, and the usual cosmic time interval dt(z). Although a redshift coordinate is used, there is no pertinent dependence on coordinate system because of the normalization in equation (5). Any observer, whatever its coordinate system, finds the dimensionless ratio of two numbers to be covariant. It is assumed that the observer can derive the temporal change in the total number of BHs, as three-space evolves, because he lives in a representative comoving volume that enjoys the same evolution in BH formation as the whole universe.

One finds numerically, using the data in [4], that Λ(z)/Λ(0) roughly scales as $(1+z)^{-0.36}$ for z<1. I.e., Λ(z) changes by less than ~30% more recently than a redshift of unity. Furthermore, a rapid decline in the dark energy density by a factor of 5 occurs from z=1-3, as one moves to before the peak in the star formation history of the universe. The value of the equation of state parameter w=p/ρ (for pressure p and density ρ) is estimated to be w= $-1.061^{+0.069}_{-0.068}$ in [5], for z<1. These data are consistent with equation (5), because a constant w= -1.1 corresponds to a change in Λ(z)/Λ(0) of about 30%, for z<1. Conversely, because the bulk of the stellar mass BHs appears not to have been formed before z=1, one expects to have w< -1 for z>1. If many (comparable to the stellar mass BH count) primordial BHs of ~$10^{15}$ g exist [3] that are evaporating for z<3, then values of w larger than -1 are possible during that cosmic epoch.

**Acknowledgements:** The author is grateful to Elly van 't Hof and Aycin Aykutalp for discussions on space-time topology and black hole formation.